# Angle-dependent resonant dynamics of stripes and skyrmions in Re/Co/Pt multilayers

Sukanta Kumar Jena[1,2,3,4,*], Kilian Lenz[3], Mateusz Zelent[5#], Mathieu Moalic[5], Artem Lynnyk[1], Aleksiej Pietruczik[1], Paweł Aleszkiewicz[1], Ewelina Milińska[1], Jürgen Lindner[3], Andrzej Wawro[1]

1. Institute of Physics, Polish Academy of Sciences, Aleja Lotników 32/46, 02-668 Warsaw, Poland
2. Jerzy Haber Institute of Catalysis and Surface Chemistry, Polish Academy of Sciences, Niezapominajek 8, 30-239 Krakow, Poland
3. Institute of Ion Beam Physics and Materials Research, Helmholtz-Zentrum Dresden-Rossendorf, Bautzner Landstraße 400, 01328 Dresden, Germany
4. Bilakankuda, Kapileswarpur, Puri, Odisha 752013, India
5. Institute of Spintronics and Quantum Information, Faculty of Physics and Astronomy, Adam Mickiewicz University, Uniwersytetu Poznańskiego 2, 61-614 Poznań, Poland

*sukaphysics@gmail.com
# mateusz.zelent@amu.edu.pl (micromagnetic simulations)

**Abstract:**
	The dynamic behavior and stabilization of skyrmions in magnetic multilayers are critical for advancing spintronic and magnonic technologies. In our study, we investigate the static and dynamic properties of [Re/Co($d_{Co}$)/Pt]$_{20}$ multilayers with varying Co thicknesses (6–24 Å), showcasing a transition from out-of-plane to in-plane magnetic anisotropy. Magnetization reversal leads to a transformation from labyrinth domains to skyrmion bubbles due to the interfacial Dzyaloshinskii-Moriya interaction (iDMI). Using angle-dependent imaging at remanence, we confirm that skyrmions can be stabilized without an external magnetic field at specific polar angles, with the stabilization angle increasing alongside Co thickness. Ferromagnetic resonance spectroscopy reveals four distinct resonant modes, including low-frequency (2–18 GHz), high-frequency (20–35 GHz) modes, depending on the magnetization texture. The frequency range of these modes narrows with decreasing effective anisotropy and iDMI strength decreases in thicker Co layers in perpendicular configurations. Moreover, we observe a decrease in effective Gilbert damping with increasing Co thickness, highlighting the potential for efficient energy dissipation. These findings link between material properties and skyrmion dynamics directly and demonstrate tunable resonant modes for magnonic devices. By addressing both static and dynamic aspects, our work advances the development of next-generation spintronic and magnonic applications.

**Introduction:**
	The presence of asymmetrical heavy metal layers adjacent to Co leads to an asymmetric interaction known as the interfacial Dzyaloshinskii-Morya interaction (iDMI)[1]. Skyrmions are topologically stable chiral quasi-particles that can be visualized in systems with lack of inversion symmetry arising at the interface of ferromagnetic and heavy metal heterostructures, i.e. with iDMI and high spin orbit coupling[2,3,4,5]. Skyrmions could be potentially used in magnetic data storage devices such as racetrack memory [6,7], or computational devices such as skyrmion-based shuffle computing [8], probabilistic computing [9], skyrmionic magnetic tunnel junctions[10], reservoir computing and other[11,12]. In the last few years, researchers have been working on the fast data processing with low power consumptions. A lot of research focusses on the stabilization of skyrmions at room temperature in thin film heterostructures[13] and on the static characterization of skyrmions in thin-films[1].
	The stabilization of skyrmion in absence of any external magnetic field is important for the data storage technology such as racetrack memory[14]. In this light, it was reported that the skyrmion stabilization at remanence could be possible by applying the magnetic pulses or simultaneously applying the static magnetic field in the both in-plane (IP) and out-of-plane (OOP) direction[15]. Salikhov *et al.* demonstrated that after reducing the applied tilted static field to zero at a particular angle is sufficient to stabilise the bubble at remanence in Pt/Co/Pt multilayers with high perpendicular magnetic anisotropy (PMA)[16].

Recently, the dynamic nature of skyrmions was investigated by applying GHz frequencies in the presence of a magnetic field. Satywali *et al.* studied the magnetization dynamics in Ir/Fe/Co/Pt heterostructures with overall IP anisotropy confirmed by FMR[17]. It was observed that there are two distinct modes, a low frequency (LF) mode and a high frequency (HF) mode with negative dispersion before the saturation. The LF mode and HF mode are contributed to the counter clock wise (CCW) gyration of the skyrmion cores and the CCW precession modes, respectively. The modes are highly sensitive to the dipolar interaction between the two ferromagnetic layers. Similar observations have been found in the Pt/CoFeB/AlOx heterostructures with overall IP anisotropy studied by Srivastava *et al.*[18]. This investigation revealed the existence of three distinct resonant modes such as the LF mode, intermediate frequency (IF) modes, and the HF mode, respectively. The LF mode was attributed to the uniform precession of the background magnetization between the skyrmion and the uniform precession of the background magnetization on the top surface of the sample. However, the precession of skyrmion edges in the central part of the sample is responsible for the IF modes[18,19]. Moreover, the in-phase precession of the reversed magnetization within the skyrmion cores is responsible for the HF mode and its corresponding excitations extends from the bottom up to the top layers[18]. The above-mentioned properties make skyrmions suitable candidates for usage in rf oscillators[20], rf sensors[21], and reconfigurable magnonic crystals for magnonic devices[22]. Theoretical studies[23] reveal the rich variety of skyrmion eigenmodes in presence of a radio frequency, whereas the azimuthal bound states and the breathing modes are observed in bulk crystals with low damping material at low temperatures[24]. There are few reports dedicated to the dynamics of skyrmions and chiral stripes. An important aspect of chiral stripes is that exciting them with GHz frequencies could be beneficial for magnonic spintronics devices[25,26]. Recently, it has been found that parallel stripes can behave as reconfigurable magnonic devices[27] possessing a higher group velocity of 2.4 km/s as compared to the velocity of skyrmions and the domain wall. However, the presence of iDMI in PMA heterostructures has not yet been studied by broadband FMR. Also, the transition of resonant modes from a higher frequency range to low frequency modes by decreasing the effective anisotropy constant was not studied so far.

In this paper, we explore the stabilization of magnetic skyrmions by applying a magnetic field and, further the stabilization of a skyrmion lattice from labyrinth and strip domains at remanence by applying an inclined field. We measured the corresponding resonant modes by FMR up to 35 GHz with the static magnetic field parallel and perpendicular to the sample surface. Further, we performed micromagnetic simulations to understand the characteristic and localization of modes arising from the skyrmions and chiral stripes.

**Static magnetic properties:**

**The magnetization reversal of Re/Co/Pt multilayers:**
We measure the saturation magnetization values of the [Re (10 Å)/Co($d_{Co}$)/Pt (10 Å)]$_{20}$ multilayers for IP and OOP direction by SQUID magnetometry at room temperature. The values are listed in Table 1 in the methods section. We found a slightly higher saturation magnetization $M_s$ compared to the Co bulk value of 1440 kA/m. This might arise from the magnetic proximity effect at each Co/Pt interface adding some parallel magnetic moment to the whole heterostructures[28,29]. The magnetization reversal occurs from negative saturation field to positive and vice versa. At remanence, the total magnetic moment nearly equals zero suggesting that the magnetic moment in upward and downward directions are equal. Consequently, the formation of labyrinth domains occurs at remanence.

Here, we calculate the effective iDMI amplitude $D_{eff}$ using the $K_{eff}$-method described in reference[30]. The exchange constant $A_{ex}$ equals 22, 24, and 28 pJ/m for Co thicknesses of 8, 10, and 12 Å, respectively, and is taken from literature[30,31,32]. In the epitaxial samples, $A_{ex}$ is substantial higher than the sputter-grown magnetic thin films with lower thicknesses[30,31,32]. Interestingly, it has been observed that the value of $D_{eff}$ decreases with increasing Co thickness. However, the surface iDMI constant $D_s$, which is independent of the thickness[33], remains constant throughout all thicknesses of Co as can be seen from table 1.

**Table 1. List of magnetic parameters derived from SQUID measurement.** Magnetic properties of of the [Re (10 Å)/Co($d_{Co}$)/Pt (10 Å)]$_{20}$ multilayers: demagnetization energy $\mu_0 M_s^2/2$, effective anisotropy $K_{eff}$, IP uniaxial anisotropy $K_U$, effective anisotropy field $\mu_0 H_{Keff}$, OOP and IP saturation field $\mu_0 H_{sat\,(oop/ip)}$, volume anisotropy $K_V$, surface anisotropy $K_S$, domain width parallel to the stripes after demagnetization $W$, domain wall width $\Delta$, calculated effective iDMI strength $D_{eff,Keff}$, using the exchange constants $A_{ex} = 22, 24,$ and $28 \frac{pJ}{m}$, and the surface iDMI parameter $D_{s,Keff}$. $SK_{n,m}$ is the maximum number of skyrmions.

| Sample | $M_s$ (kA/m) | $\frac{\mu_0 M_s^2}{2}$ (MJ/m$^3$) | $K_{eff}$ (MJ/m$^3$) | $K_u$ (MJ/m$^3$) | $\mu_0 H_{Keff}$ (T) | $K_V$ (MJ/m$^3$) | $K_S$ (mJ/m$^2$) | $W_{stripes}$ (nm) | $D_{eff,Keff}$ (mJ/m$^2$) | $D_s$ (pJ/m) | $SK_{n,m}/\mu m^2$ |
|---|---|---|---|---|---|---|---|---|---|---|---|
| S1 | 1380 | 1.19 | -0.49 | 1.69 | -0.71 | 2.2 | 0.56 | 86 | 3.72 | 2.99 | 18 |
| S2 | 1486 | 1.38 | -0.23 | 1.62 | -0.32 | | | 84 | 2.98 | 2.98 | 17 |
| S3 | 1441 | 1.30 | -0.07 | 1.37 | -0.09 | | | 85 | 2.47 | 2.97 | 13 |
| S4 | 1435 | 1.29 | 0.15 | 1.14 | 0.20 | | | 91 | | | 12 |
| S5 | 1447 | 1.31 | 0.32 | 0.99 | 0.45 | | | 90 | | | 1 |
| S6 | 1453 | 1.32 | 0.41 | 0.91 | 0.57 | | | 93 | | | 8 |

**Field dependence of the magnetic domain evolution and observation of skyrmions in Re/Co/Pt multilayers:**

To observe the influence of the magnetic field on the domain evolution, we took magnetic force microscopy (MFM) images in field dependence for samples S1, S2, and S3 with $K_{eff}$< 0 (OOP easy axis) as estimated from the M-H-loop. We recorded the field-dependent magnetic domains with increasing the magnetic field[19]. Remarkably, we observed that with increasing the magnetic field in the positive direction, the labyrinth domains transform into bubble skyrmions and eventually reach a saturation state for sample S1 as shown in Figs 1(a)-1(d). Similarly, the transition from labyrinth domains to skyrmion is observed as the magnetic field increases in positive direction, for sample S2 and S3, as shown in Figs. 1(e-h) and 1(i-l), respectively. The skyrmion bubble density increases significantly with decreasing uniaxial anisotropy energy from 1.69 MJ/m$^3$ to 1.37 MJ/m$^3$ with increasing the thickness. At an external applied field of $\mu_0 H = 0.12$ T, three different states, i.e. the spin-polarized state, the skyrmion state, and the skyrmion lattice state, are observed for samples S1, S2, and S3, respectively, as shown in Figs. 1(d, h, l). The formation of a bubble skyrmion lattice is due to the balance of uniaxial anisotropy and iDMI energy. From the above discussion, we conclude that the evolution of the magnetic domain under the influence of an external magnetic field for samples with negative effective anisotropy (OOP easy axis) exhibits a transition from labyrinth domains to skyrmion structures as the external field increases providing a relationship between skyrmion density and effective anisotropy.

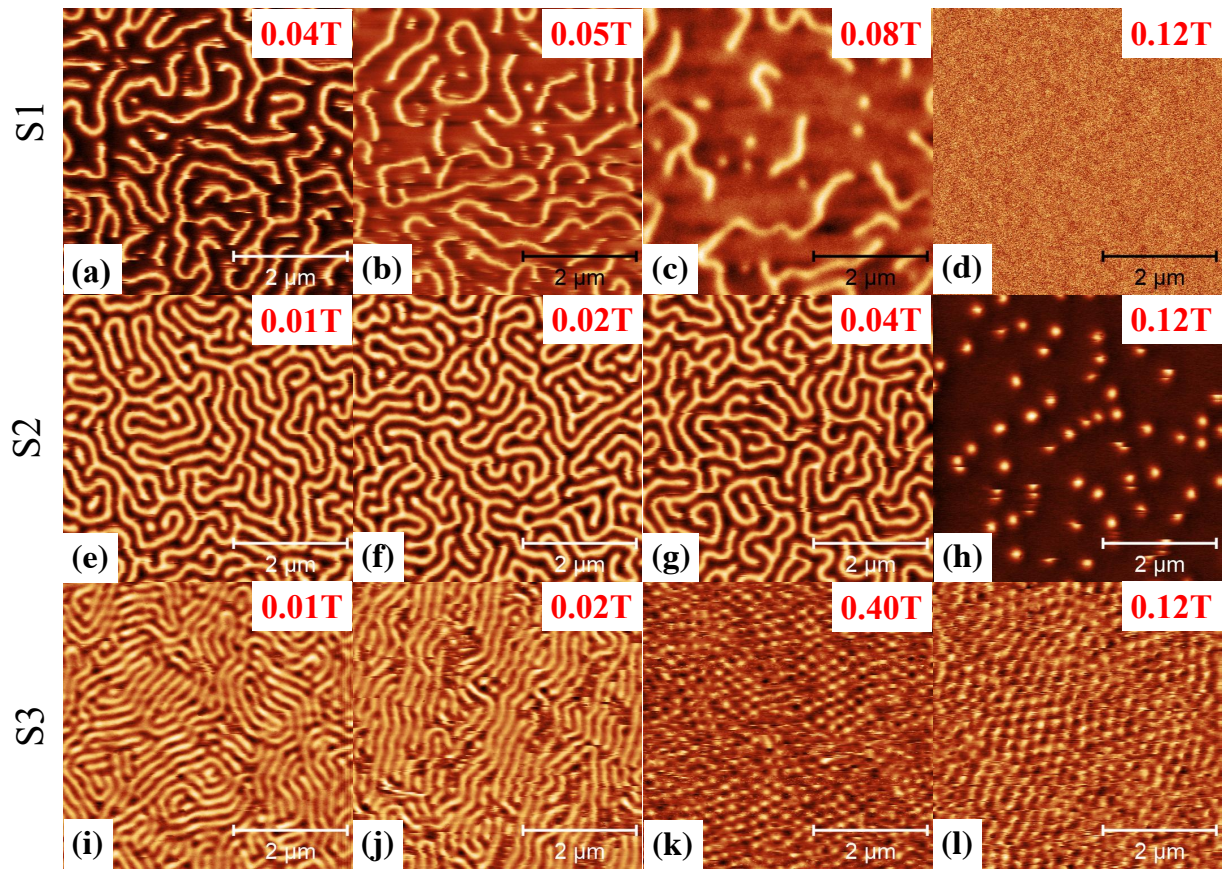

Figure1: *Evolution of the magnetic domains under applied field for Co thicknesses of 8 Å, 10 Å, and 12 Å (top to bottom row, samples S1-S3, respectively). All MFM images have the same field-of-view equal to* $25\ \mu m^2$.

**Stabilization of skyrmions at remanence:**

In the previous section, we observed that the skyrmions are stabilized by applying the magnetic field normal to the sample. By decreasing the field further to zero, the skyrmion state vanishes and the labyrinth state is restored. In this section, we investigate how to stabilize the magnetic skyrmions at remanence by using the angle-field-dependent MFM method for both IP and OOP easy-axis samples as described in Ref.[16]. Here, we use the following protocol as described in the method section.

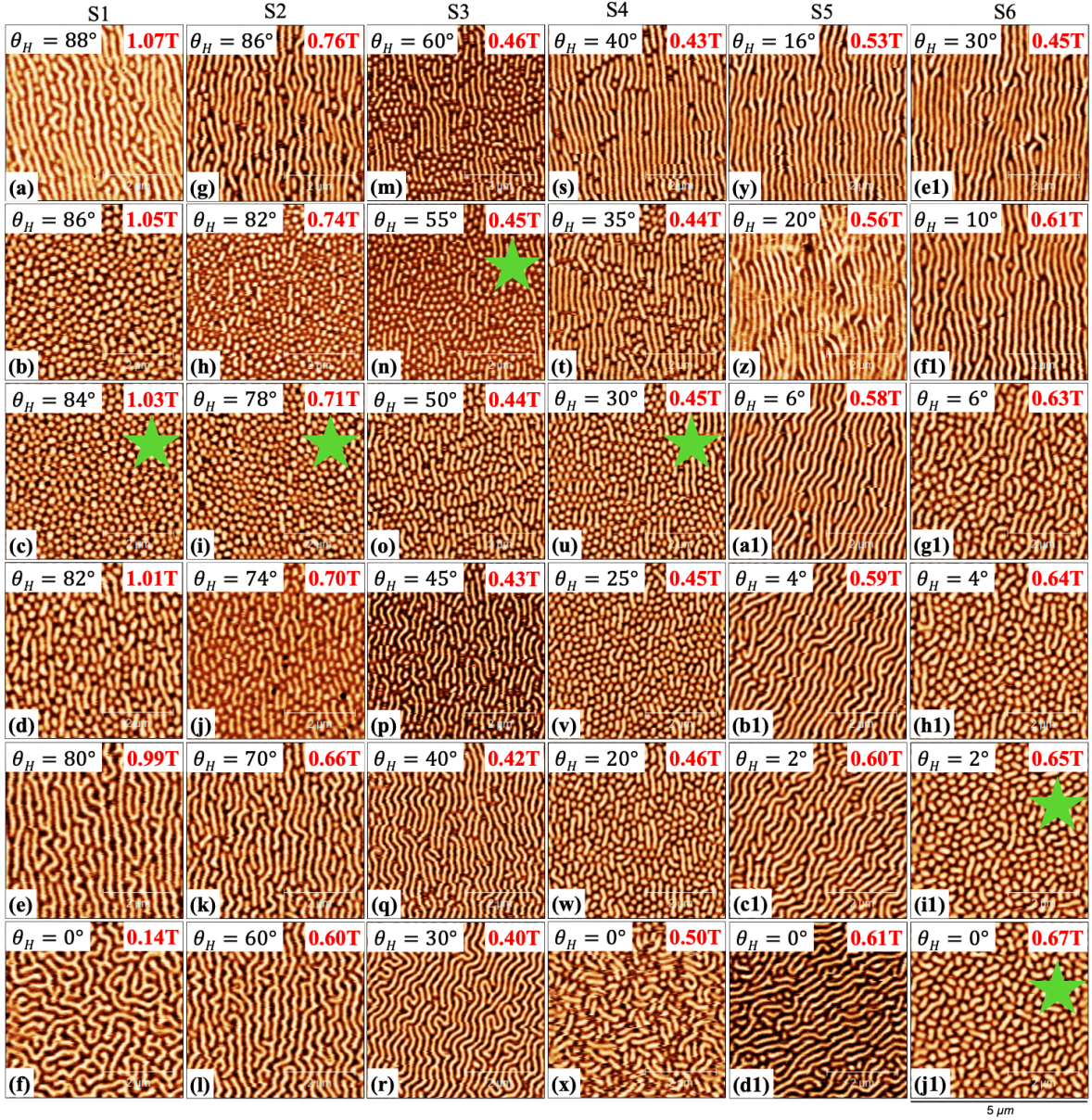

Figure 2: (a-f), (g-l), (m-r), (s-x),(y-d1) and (e1-j1) represent the angle-dependent MFM for the Co thicknesses of 8 Å, 10 Å, 12 Å, 16 Å, 20 Å, and 24 Å, respectively, at remanence after applying different DC saturation fields under different angles and reduction to zero according to the protocol described in the main text. All MFM images have the same area-of-view equal to 25 $\mu m^2$.

The various magnetic structures are stabilized when the external positive DC saturation field $\mu_0 H_{+DC,Sat}$ is applied to the sample at different polar angles $\theta_H$ (Fig. 2). When we apply the external field in the IP direction, i.e., $\theta_H$= 90°, the labyrinth domains align in the field direction and become parallel stripes (sample S1 with 8 Å Co). The perpendicular component of the magnetization increases with increasing angle ($\theta_H$) of the applied DC field to the sample, breaking the symmetry. The symmetry breaking leads to the transformation of parallel domains into skyrmions because the tilted field provides the additional perpendicular component. Upon reducing the field strength to zero, the energy of the domains is not sufficient to revert them to their original state. Therefore, the competition between perpendicular components and the effective anisotropy field at particular angle plays an important role in the formation of the skyrmions.

When we further increase the angle, the perpendicular component becomes dominant. This leads to the breaking of the parallel stripes (Fig. 2(a)) into individual skyrmions. The skyrmion lattice is formed at $\theta_H = 86°$ as shown in Fig. 2(b). At lower angle, $\theta_H = 82°$, the perpendicular component is dominant even more and, as a result the demagnetization energy is higher in the direction normal to the sample plane. This leads to the transformation of the skyrmions into the labyrinth state with a zig-zag shape. Furthermore, we observe the absence of skyrmions at $\theta_H = 80°$ in Fig. 2(e), while a pure labyrinth state is stabilized at $\theta_H = 0°$ as shown in the Fig. 2(f). For other thicknesses of Co, the effective anisotropy $K_{\text{eff}}$ increases with the Co thickness becomes negative for sample S4. The field angle for the stabilization of the skyrmions decreases with increasing effective anisotropy energy with Co thickness. The above described phenomenon occurs for the skyrmion stabilization by tilted field angle for all thickness of Co within a wide range of effective anisotropies. The skyrmion density vs. angle of the applied field is given in Fig. 3.

Here, we have noticed that with increasing the Co thickness, the angle for skyrmion stabilization decreases from $\theta_H = 86°$ to $\theta_H = 0°$ for sample S5 (20 Å Co). We found the highest skyrmion density of 17 skyrmions/μm² for sample S1 as shown in Fig. 3. Since, the samples S2 and S4 exhibit the opposite sign of effective anisotropy strength, the maximum number of skyrmions is stabilized at angles of $\theta_H = 78°$ and $\theta_H = 30°$ for samples S2 and S4, respectively. Furthermore, increasing of the Co thickness stabilizes the maximum number of skyrmions at $\theta_H = 2°$ for sample S6 (24 Å Co). The green asterisks in Fig. 2 mark the stabilization of maximum skyrmion numbers for different samples in a particular angle at remanence.

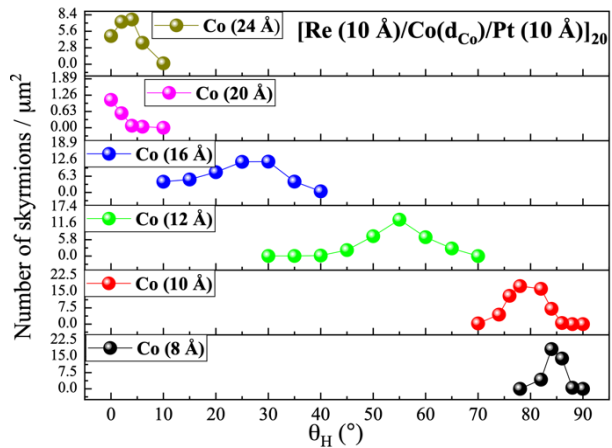

Figure 3: *Dependence of the skyrmion density in remanence on field angle and Co thickness for the [Re (10 Å)/Co($d_{Co}$)/Pt (10 Å)]$_{20}$ samples. The skyrmion density at different angles was stabilized by using protocol described in the main text.*

**Dynamic properties**
**Dynamic characterization for IP, OOP, and tilted angles**:
We performed FMR measurements for field IP and OOP configurations. The DC magnetic field was swept from negative direction to positive direction in the field range of -2 T to +2 T. We varied the microwave frequency from $f$ = 0.5 GHz to 35 GHz. Fig. 4 shows false-color plots of the FMR signal intensity with respect to field strength and frequency. Several resonant modes are observed and analyzed. We observe the spin-polarized uniform mode known as the Kittel resonance mode (denoted by KM in Fig. 4) with positive slope for both IP ($\theta_H = 90°$) and OOP ($\theta_H = 0°$) configurations for all samples after passing the saturation field. Beside these modes, we observe up to three distinct modes in the field range between negative and positive saturation. We attribute them to the dynamics of the magnetic labyrinth domains, parallel domains, and skyrmion states.

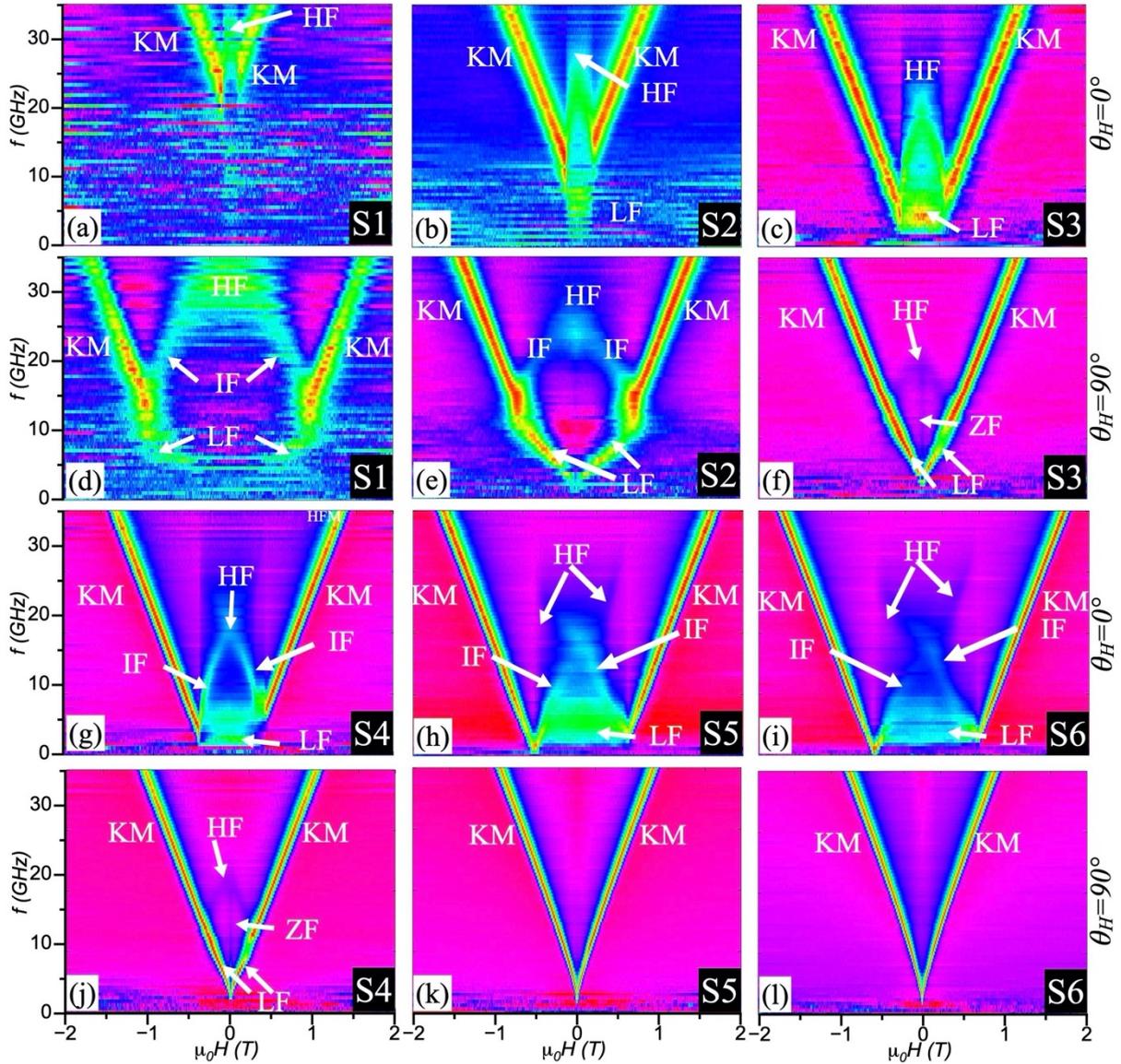

Figure 4: *(a,-c and g-i) The field-frequency-dependence of the FMR signal (color-coded) in out-of-plane configuration ($\theta_H = 0°$) of samples S1 to S6, respectively. (d-f and j-l) same for the in-plane configuration ($\theta_H = 90°$).*

For the lowest Co thickness of 8 Å (sample S1), we find the KM starting at higher frequencies around $f$ = 25 GHz for the OOP configuration and $f$ = 15 GHz for the IP configuration. Additionally, a high-frequency (HF) mode is observed above $f$ = 30 GHz close to zero field in the OOP configuration, which resembles the resonant modes arising from the presence of both labyrinth domains and skyrmions as depicted in Fig. 4(a). Additionally, three other distinct resonant modes appear, such as the HF mode starting from $f$ = 27 GHz, an intermediate-frequency (IF) mode appearing between $f$ = 8-27 GHz and a low-frequency (LF) mode in between $f$ = 5 and 10 GHz for sample S1 in IP configuration as shown in Fig. 1(d). All these modes are contributed due to the excitations of the parallel stripe domains.

When the Co thickness is increased to 10 Å (sample S2), we observe several changes in the FMR spectra showing that the starting frequency of the KM is shifted towards lower frequencies for both IP and OOP configuration. The KM is accompanied by an additional HF and LF mode. The HF mode exhibits a negative dispersion in OOP direction. For the IP configuration, the KM is accompanied by HF, IF, and LF modes for sample S2 as previously mentioned also for sample S1. However, the starting frequency of the resonant modes is shifted towards lower frequencies. Similarly, in the OOP configuration of sample S3, the Kittel mode is observed along with a HF mode and LF mode like for sample S2. However, the starting frequency for all these modes in S3 is lower as compared to sample S2. The low frequency modes are observed in non-uniform domain regions (unsaturated magnetic region) for samples S2 and S3 in OOP configuration. For the IP field configuration, sample S3, in addition to the LF mode and HF mode, exhibits another resonant mode at zero field, which is constant in field for frequencies up to 20 GHz. It starts from $f = 5$ GHz. We denote it as zero-field-frequency (ZF) mode. A similar type of ZF mode is observed for sample S4 in the IP configuration.

By further increasing the Co thickness from 16 Å to 20 Å and 24 Å (samples S4 - S6), the distinct intermediate frequency (IF) mode is clearly observed in the OOP configuration as depicted in Figs. 2(g)-2(i). Alongside the IF mode, the other modes, such as HF and LF mode are also observed, which are associated with the formation of skyrmions and the labyrinth domains as mentioned above. In the spin-polarized state, the KM is observed. With increasing the magnetic field from negative saturation towards positive saturation, the KM mode splits into the LF and IF modes where the IF mode exhibits the negative slope in frequency-field-dependence in the case of samples S4, S5, and S6. Eventually, both IF modes (in positive and negative field regions) approach each other with increasing frequency and merge at a frequency around $f = 18$ GHz. Moreover, both IF modes subsequently split then into two HF modes, which shows the non-symmetric frequency-field-dependence behavior in nature and displays a positive dispersion relation in sample S4, S5 and S6 as shown in the Fig. 2(g), (h), and (i).

In comparison to samples S5 and S6 in OOP configuration, sample S4 exhibits a narrow range spread of the HF mode in the frequency spectrum. Notably, in the previous studied multilayers Ir/Fe/Co/Pt and Pt/CoFeB/AlO$_X$, the presence of the merging IF modes and subsequent splitting to HF modes was not observed[17,18]. The broader spread of the HF modes provides a valuable insight into the existence of skyrmions over a wide range of applied magnetic fields, particularly in the intermediate unsaturated regions between the magnetic saturation points, in the case of samples exhibiting IP anisotropy[19]. For the IP FMR configurations samples S5 and S6 exhibit only the KM as illustrated in Figs. 4(k & l).

To better understand the dynamical behavior at those particular angles where the skyrmions are stabilized at remanence, we performed FMR measurements at additional field angles. The false-color plots of the FMR spectra in Fig. 5 exhibit different distinct excitation patterns compared to the IP and OOP FMR measurements shown before, where the remanent state corresponds to the skyrmion states at these particular angles for both samples S2 and S4. The FMR spectra of S2 [see Figs. 5(a-c)] shows two distinct modes in addition to the Kittel mode.

On the other hand, the sample S4 exhibits three distinct modes, including the KM at an angle of 30°. However, it does not show the intermediate frequency mode as compared to the OOP configuration for S4. The HF mode is clearly visible at $\theta_H = 30°$ as compared to $\theta_H = 90°$ (IP), whereas the ZF mode is absent for that particular angle. From the combined analysis of IP, inclined angle and OOP measurements, it is evident that the different resonant modes depend on the Co thickness and ultimately rely on the anisotropy of the system.

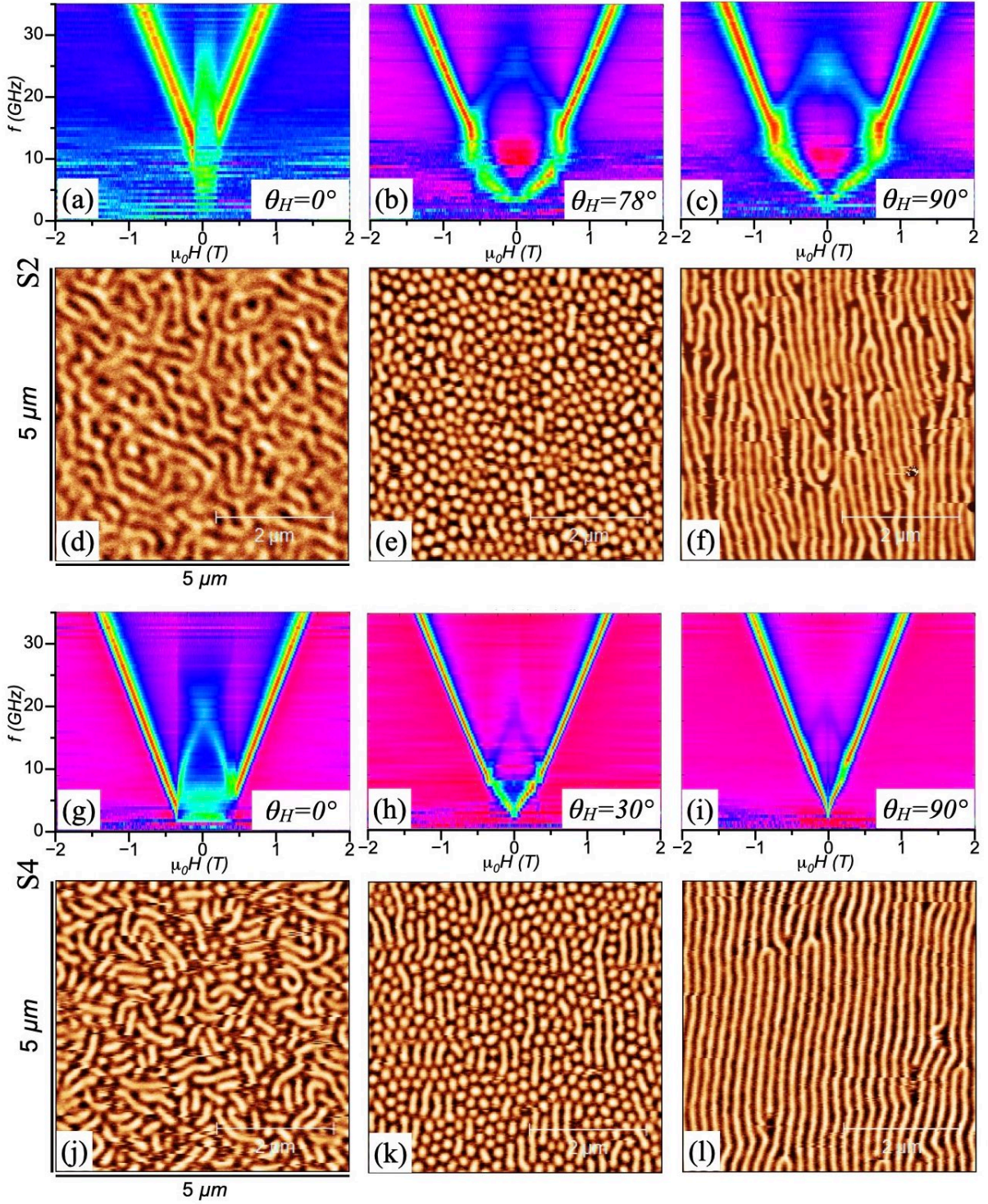

Figure 5: *(a,b,c) False-color plots of the FMR signals of sample S2 vs. field and frequency for angles θ$_H$ of 0°, 78°, and 90°, respectively. (d,e,f) Corresponding MFM images taken at remanence . (g,h,i) FMR spectra at angles θ$_H$=0°, 30°, and 90° for sample S4, which possesses in-plane anisotropy. (j,k,l) Corresponding MFM images at remanence. Note that the FMR spectra of samples S2 and S4 at 0° and 90°are repeated from Fig. 4 for comparison.*

**Micromagnetic simulations:**

To corroborate the experimental results, we performed micromagnetic simulations for both statics and dynamics. The field-dependent domain evolution, including the skyrmion formation is simulated using our own fork of Mumax3[34] called Amumax [https://zenodo.org/records/13349742]. The simulations were performed for a finite 2μm × 2μm film with periodic boundary conditions along the OOP axis with thirty-two iterations in both negative and positive z-directions and full model along the thickness including magnetic and non-magnetic layers throughout, where the cell thickness of 1.0 or 1.6 nm corresponds to the thickness of the magnetic layer in the real sample S2 or S4, respectively. Details of the micromagnetic simulations are given in the Methods section.

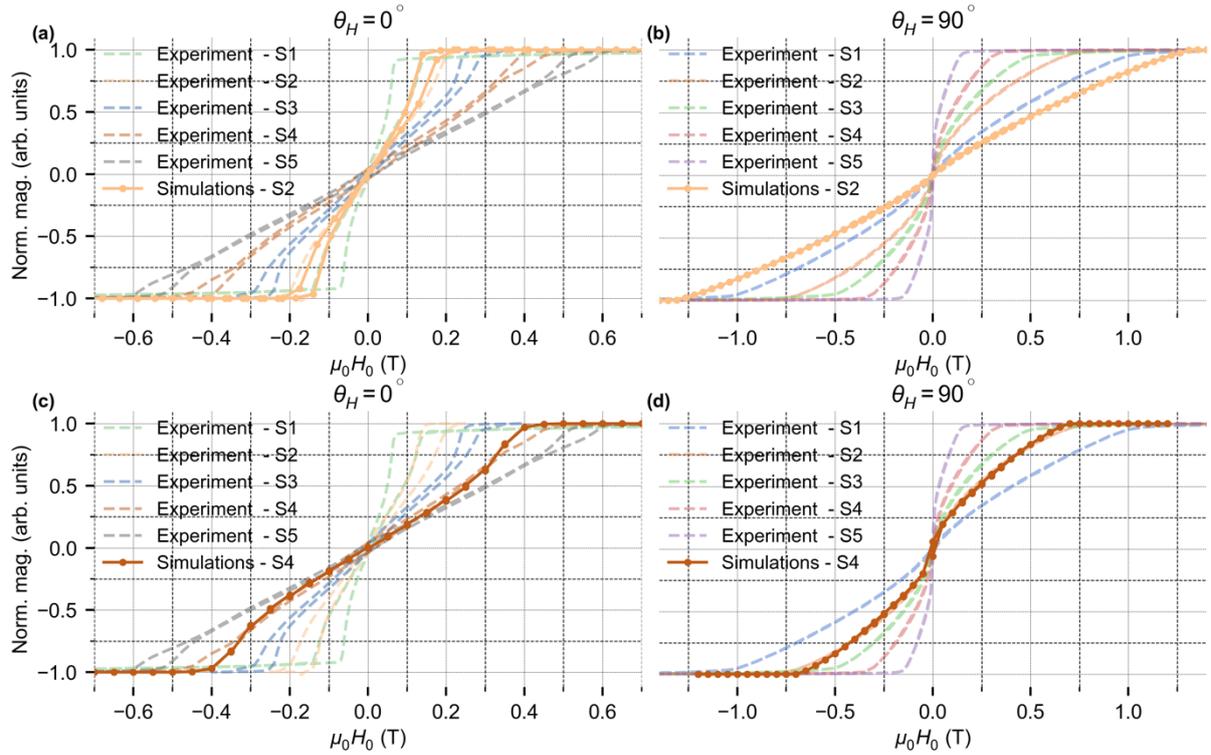

Figure 6: *Hysteresis loops comparing simulation results and experimental measurements (a,b) sample S2 for out-of-plane and in-plane field, respectively. (c,d) same for sample S4 Each panel shows the normalized magnetization ($M/M_s$) as a function of the applied magnetic field. Panels The experimental data is represented by dashed lines and the corresponding simulation results are shown by solid lines with circles.*

In thin multilayer structures, the ratio of the widths of domains with magnetization oriented downward to those oriented upward is primarily influenced by the Zeeman energy. The size and morphology of these domains are determined by a combination of anisotropy and domain wall energy, while their chirality is governed by the iDMI. We performed micromagnetic simulations of the field-dependent hysteresis loops using various sets of experimentally measured parameters, fine-tuning these parameters to achieve the closest alignment with the experimental observations through a comprehensive 3D numerical model. The simulation results for samples S2 and S4 (using the parameters given in the Methods section) are presented in Figs. 6(a-d) as solid lines. Our numerical results show strong qualitative agreement with the experimental data, represented by dashed lines. The primary criterion for parameter selection was to accurately reproduce the shape of the hysteresis loop in the perpendicular magnetic field configuration ($\theta_H = 90°$) for S2 and S4 independently. For both samples, S2 and S4, the selected parameter sets resulted in IP hysteresis loops with a steeper slope compared to the experimental results. However, the overall magnetization evolution of the samples was accurately reproduced. Notably, for sample S4 in the IP configuration, the shape of the hysteresis loop closely aligns with that of sample S2, particularly demonstrating a strong match in the OOP loop. These findings indicate that the chosen parameters effectively replicate the experimental measurements qualitatively. The observed discrepancies can likely be attributed to the fact that the simulations were conducted at 0 K, thereby neglecting thermal fluctuations, as well as potential differences in effective and surface anisotropies that may be present in the experimental samples but were not accounted for in the model.

In the simulations, during gradually decreasing the amplitude of external magnetic field a labyrinthine domain pattern emerges at zero magnetic field. As the magnetic field strength increases, these labyrinthine domains gradually transform into isolated skyrmions. A noteworthy attribute of the intricate domain and skyrmion configurations is the pronounced variation in their configuration along the thickness of the multilayer (see Sup. Fig. SM-5). This variation in core structure is attributed to the fact that the [Re (10 Å)/Co($d_{Co}$)/Pt (10 Å)]$_{20}$ multilayers layers are coupled primarily through dipolar interactions. While these interactions are strong enough to ensure the alignment of the core centers across the layers, they are insufficient to enforce a coherent magnetization profile, resulting in skyrmion cores and domain structures that are not identical across the thickness. The skyrmions in the bottom half of the stack remain topological (Q=1), while the top half comprises largely non-topological bubbles, with multiple chiral kinks (Q>1).

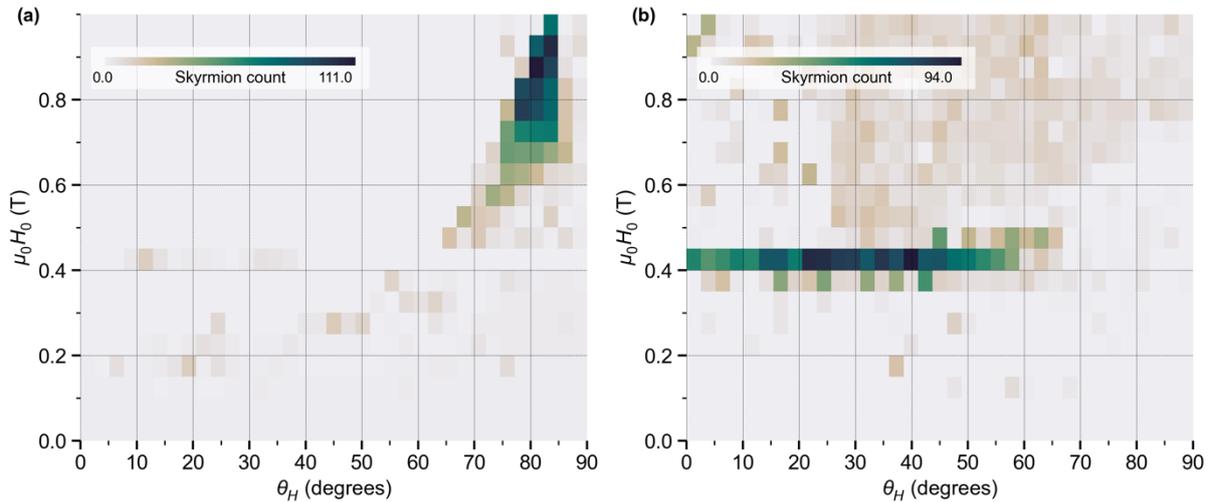

Figure 7: *The skyrmion density maps for samples S2 and S4 as a function of the external magnetic field strength and the field angle relative to the sample plane ($\theta_H$).*

Fig. 7 presents the skyrmion density maps at remanence for samples S2 and S4 as a function of the external magnetic field strength and the field angle relative to the sample plane ($\theta_H$). The simulations commenced with the labyrinthine domain pattern as initial state, uniformly across all cases. Subsequently, an external magnetic field of specified amplitude and angle ($\theta_H$) was applied. The system was allowed to reach equilibrium under the tilted magnetic field, then the structure was further relaxed into its remanent state. In this remanent state, the number of circular domains (bubbles) were counted and classified as skyrmions. This result demonstrates the influence of both field intensity and direction on the stabilization and density of skyrmions in multilayers. Fig. 7(a) shows the skyrmion density map for sample S2. The maximum skyrmion density is observed near the saturation field, as indicated by the dark regions in the color map. The highest number of the skyrmions, 121 in a 4 µm² area (which corresponds to 30 skyrmions on 1 µm²), has been observed for an applied field of 1.2 T at an angle of $\theta_H \approx 80°$. These results suggest that a critical factor in the formation of skyrmions in S2 geometries is the high angle of the external magnetic field, approaching almost IP orientation. Panel 7(b) illustrates the skyrmion density for sample S4. In this case, the skyrmion density peaks at lower field strengths compared to S2, reflecting differences in anisotropy and iDMI strength due to the increased Co thickness. The map reveals a narrower range of field angles where skyrmions are stabilized, consistent with the reduction in effective anisotropy. The highest skyrmion count, namely 94 on a 4 µm² area (equivalent to 24 skyrmions per 1 µm²), was observed at 0.45 T and $\theta_H \approx 40°$. These findings indicate that for S4 geometries, the high amplitude of the external magnetic field, approaching the saturation field, is a crucial factor in skyrmion formation. The skyrmion density slightly deviates from experimental measurements as shown in the Table-I, which may be attributed to the assumed magnetic parameters and resulting minor differences in skyrmion size, as well as the method of skyrmion counting and identification. Nonetheless, for the selected material parameters, we achieved strong agreement regarding the magnetic field angle that yields the highest skyrmion density in both S2 and S4 samples. Both maps underscore the critical dependence of skyrmion stability on the interplay between field direction, field strength, and material parameters such as Co thickness and iDMI strength in skyrmion formation process, the elongation of magnetic domains, driven by the strong magnetic field along the plane of the sample, plays a pivotal role. As the field amplitude approaches the saturation field, the domains become fragmented or vanished. The residual domains that persist randomly in certain layers of the multilayer after the magnetic field is reduced, serve as nucleation sites for skyrmions.

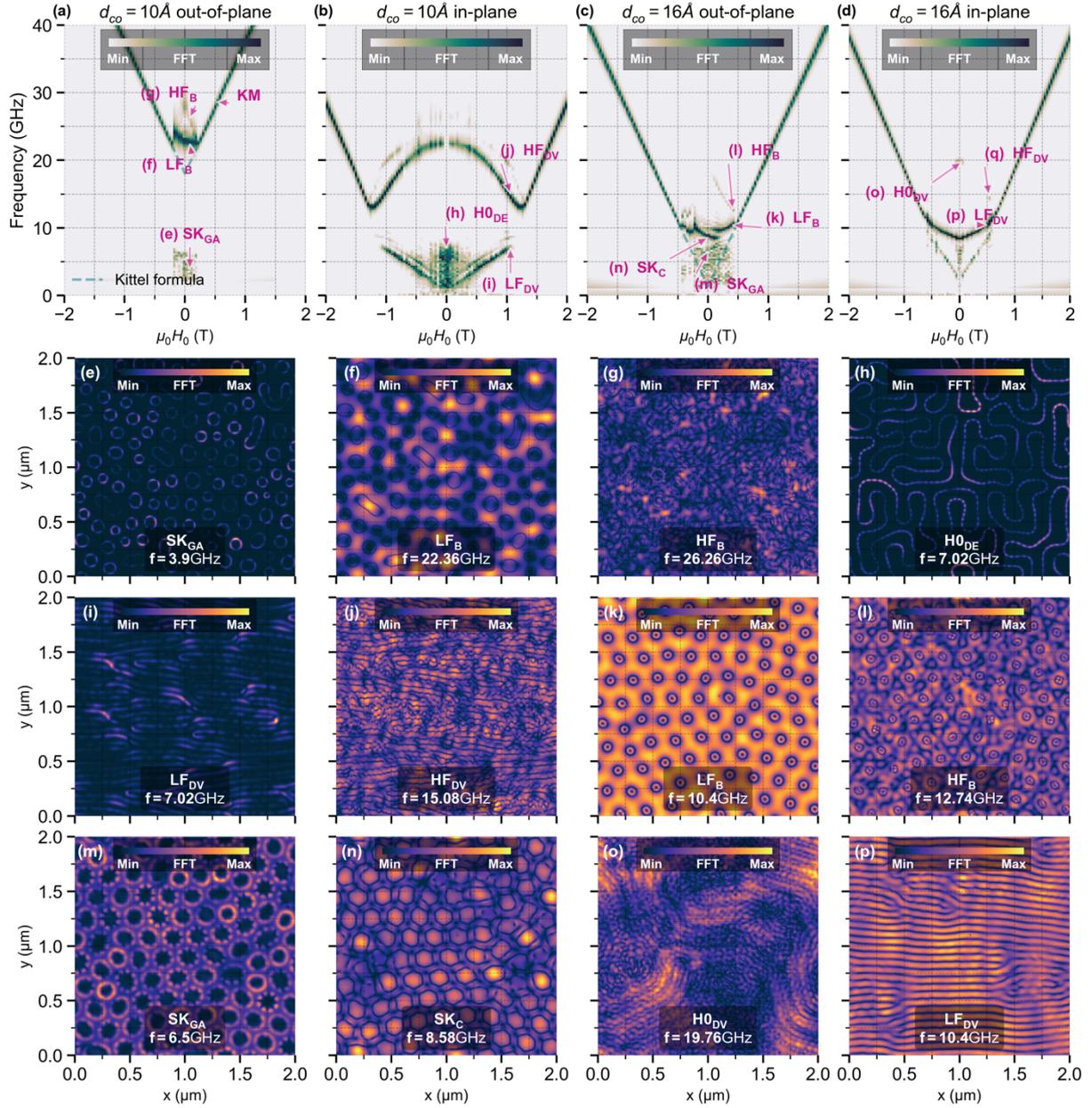

Figure 8: *(a-d) Simulated FMR spectra of samples S2 (a,b) and S4 (c,d) as a function of external magnetic field for out-of-plane and in-plane configurations. Green markers indicate the specific field and frequency values corresponding to the visualized magnetization modes shown below. (e-l) Corresponding magnetization mode maps at selected frequencies for: (e-k) out-of-plane and (l) in-plane configurations.*

Next, we discuss the dynamical response of the system, where the FMR is computed under different applied fields, as shown in Figs. 8 (a-d) for both the S2 and S4 sample in IP and OOP, respectively. In Fig. 8(a), the evolution of the FMR spectrum for the S2 sample in OOP configuration is presented. Two distinct field ranges can be identified, each corresponding to different magnetic texture responses to the external magnetic field excitation.

In the first range, corresponding to the skyrmion phase where the magnetization texture is nonuniform, three distinct types of modes can be observed within the field range of $-0.240\,\text{T} < \mu_0 H_z < 0.24\,\text{T}$. Within this range, we identify the low-frequency bulk mode (LF$_B$) and the high-frequency bulk mode (HF$_B$), which are present between 22 GHz and 28 GHz.

These modes appear in experimental measurements for the S2-S6 samples; however, we observe a shift with respect to frequency and field in the simulations.

The amplitude of the $LF_B$ mode is primarily localized in the bulk regions between skyrmions and chiral bubbles. Due to the non-uniform magnetization profile across the thickness of the chiral bubbles, the amplitude of the mode is localized differently within the layers, i.e., concentrated inside the bubbles in some regions, while primarily situated within the bulk for others. We do not observe a hybridization with skyrmion modes. Between 2 and 7 GHz, we find multiple skyrmion resonances. Exemplarily, skyrmion gyrotropic and higher-order azimuthal modes with a frequency of 3.9 GHz are present in Fig. 8(e) and labelled as $SK_{GA}$. The frequencies of these modes vary due to the order of the azimuthal mods as well as the irregular spatial distribution (mode location) resulting from the varying texture of the domain walls and shape of the skyrmions. The blue dashed line presents solution of the Kittel formula for the homogeneous mode, including OOP uniaxial anisotropy. Along with the analytical solution in the saturation regime the resonance frequency increases with the applied magnetic field, displaying a typical V-shaped dispersion characteristic for thin films (labelled as KM). However, we have not observed strong skyrmion core resonance for S2 sample with easy out of plane. The $HF_B$ mode, as shown in Fig. 8 (g), represents a high-frequency excitation chaotically localized around the skyrmion lattice. Its amplitude is distributed irregularly across the IP spatial domain, reflecting its complex and non-uniform character. This lack of localization and irregular spatial distribution result in a lower overall amplitude compared to more coherent modes.

Simulations for the configuration with the external magnetic field applied in the plane of the sample were performed starting from a relaxed labyrinth domain state at remanence, without skyrmion states [Fig. 8(b)]. Near zero field, we can distinguish a band of modes localized along the domain walls, labelled as $H0_{DE}$ [zero-field domain edge in Fig. 8(b)]. These modes show different degrees of quantization along the domain walls and are observed in the frequency range from 0.5 GHz to 7.5 GHz. The dynamic magnetization landscape of this 7.02 GHz mode is shown in Fig. 8(h) at zero field. These modes have a nearly homogeneous amplitude distribution throughout the thickness. The second branch of modes, called high-frequency domain volume ($HF_{DV}$) modes is observed both in the domain's volume and at the domain surface. It exhibits negative dispersion, which is associated with the softening of these modes [Figs. 8(b & j)]. These modes are characterized by a pronounced asymmetry in amplitude distribution across the thickness of the sample, with strong localization in both the upper and lower layers. The upper layers display a generally higher amplitude than the lower layers–a behavior attributed to the presence of iDMI in the sample and the inhomogeneous magnetic texture through the thickness (see sup. Fig. SM-5). Mode softening is a phenomenon typically observed in systems with magnetization gradients, such as domain walls, vortices, and skyrmions[35] In these regions, the effective magnetic field and energy are reduced due to the balancing effects of exchange and anisotropy contributions, resulting in a lower energy requirement for the magnetic moments to undergo precession.

The last group of modes, labelled as low-frequency domain wall $LF_{DV}$ modes, are localized within the walls of magnetic domains. These modes form a V-shaped band in the same frequency range as the $H0_{DE}$ modes. Above 1 T, the domain walls disappear, resulting in the vanishing of this band. At higher fields, a uniform mode (KM) is observed. The $LF_{DV}$ modes are not uniform across the plane of the sample; instead, they exhibit amplitude localization in specific segments of the domain walls that are oriented at an angle to the external magnetic field. Unlike the $LF_{DV}$ modes, the $HF_{DV}$ modes do not display asymmetry or inhomogeneity along the thickness [see Figs. 9(b & f)]. The mode profile 15.08 GHz mode at $\mu_0 H_z = 0.89$ T is depicted in Fig. 8(j).

Simulations for the sample S4 exhibit a similar field dependence, with a noticeably smaller frequency difference between the bulk and skyrmion modes compared to S2. The first low frequency volume mode, referred to as (LF$_B$), is illustrated for a field of -0.37 T and a frequency of 10.4 GHz in Fig. 8(k). This mode is characterized by a high amplitude localized in the bulk around the skyrmion lattice hybridized with the first-order skyrmion radial mode. It has a uniform profile through the thickness, although the amplitude slightly varies across different layers [see Figs. 9(c & g)]. This mode displays an almost flat field dependence and is observed in the field range from -0.4 to 0.4 T. Below this mode, we observe a mode called skyrmion edge mode (SK$_{GA}$) presented in Fig. 8(m) at a frequency of 6.5 GHz and field of -0.368 T. This skyrmion mode is localized in the skyrmion edge, exhibiting an azimuthal character with varying degrees of quantization along the domain wall. This mode also shows a flat field dependence and does not exhibit significant inhomogeneity across the thickness or within the plane of the sample. Below the LF$_B$ mode, in the frequency range above 0.5 GHz to 8.5 GHz, we observed a series of degenerated skyrmion modes with different frequencies and quantization. These modes are non-uniform both in the plane of the sample and along its thickness. An example of such a mode, with a frequency of 8.58 GHz, is presented in Fig. 8(n) at zero field. This mode, observed at zero magnetic field, is a skyrmion mode with amplitude localized in the skyrmion cores (SK$_C$). This kind skyrmion core resonant modes are observed in the frequency range (7-8 GHz) in IP easy axis multilayers of Ir/Fe/Co/Pt[17]. However, Srivastava *et al.* observed this mode in the higher frequency range (12-15 GHz) in Ta/CoFeB/AlOx multilayers[18].

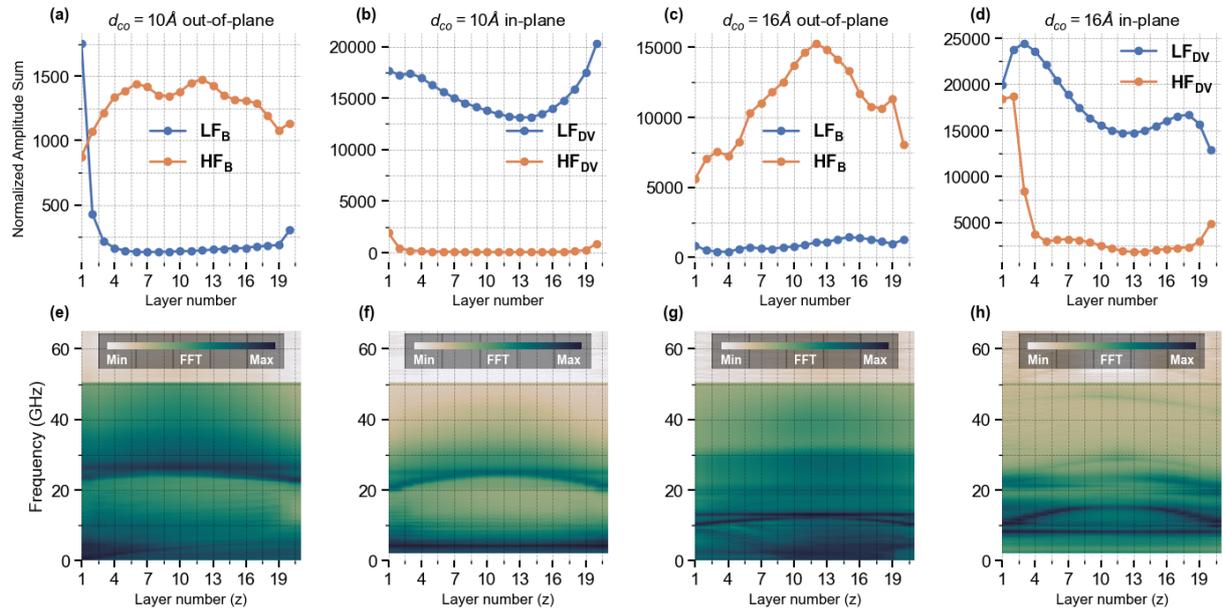

Figure 9: *Amplitude distribution across each layer for different Co thicknesses ($d_{Co}=$ 10 Å and 16 Å) and magnetic field orientations, highlighting two selected modes, i.e. the low-frequency ($L_F$) and high-frequency ($H_F$) mode.*

Simulations of the S4 system for the IP field configuration reveal a significant reduction in the intensity of the HF$_{DV}$ mode. This mode, observed at a frequency of 20 GHz at remanence, exhibits negative dispersion and decreases its frequency to 10 GHz until saturation at approximately 0.6 T. Conversely, the LF$_{DW}$ mode demonstrates two primary branches: one at higher frequency with relatively flat dependence and another at lower frequency with reduced intensity, resembling a V-shaped dependence typical of the Kittel mode. The profiles of these modes do not differ significantly and represent volume modes with amplitude localized within the domain walls, exhibiting a non-uniform spatial distribution (illustrated in Fig. 8(o) and Fig. 8(p) for the HF$_{DV}$ and LF$_{DV}$ modes at frequencies of 19.76 GHz and 10.4 GHz, respectively). The LF$_{DV}$ mode shows asymmetric amplitude localization with stronger intensity near the lower surface of the sample. In contrast, the HF$_{DV}$ mode also displays pronounced symmetry, with significant amplitude localization at the lower surface. We observe similar tendencies in experiment between S1 and S5, where also decay of HF$_{DV}$ is present. Like in the S2 sample, the LF$_{DV}$ mode is localized within the domain walls in regions perpendicular to the direction of the magnetic field as observed for parallel stripes in LSMO thin films[27]. This mode displays also a non-uniform amplitude distribution across the thickness with the maximum amplitude at both surfaces.

The quantitative difference in mode frequencies as a function of magnetic field between simulations and experiments is due to the differences in experimental parameters and the significantly higher IP anisotropy in micromagnetic simulations. Nevertheless, the experimental results were qualitatively reproduced in the numerical calculations.

**Discussion:**

Recently, it was shown that two distinct modes such LF and HF modes are observed in the frequency range of 6-12 GHz for Ir/Fe/Co/Pt, where the magnetic thickness was 10 Å and the iDMI value was about 1.4 mJ/m$^2$ ($D_S$=1.4 pJ/m)[17]. Conversely, in our Re/Co/Pt system, we have identified that the frequency range for the LF mode is $f$ = 6-10 GHz and $f$= 20-27 GHz for the HF mode, originating from a similar thickness magnetic material but having a twice larger iDMI value of 2.98 mJ/m$^2$ ($D_s$ = 2.98 pJ/m). In the micromagnetic simulations, we have observed three different modes: SK$_{GA}$, LF$_B$, and HF$_B$ for perpendicular orientation. The LF$_B$ mode is localized in between chiral bubbles and skyrmions. However, Satwyali *et al.* observed that the skyrmion core Eigen-mode is responsible for the origin of the low frequency mode in the $f$=7-8 GHz range. We have not found any skyrmion core resonant modes for sample S2. Further, we have observed the other modes called skyrmion resonance modes where spectral intensity is localized at edge of the skyrmions. Please note that Ir/Fe/Co/Pt[17] and Re/Co(10 Å)/Pt multilayers exhibit negative and positive values of anisotropy, respectively, and different values of the iDMI strength. Therefore, there is discrepancy of different observed eigen-resonant modes.

Similarly, Srivastava *et al.* investigated the system of Ta/CoFeB (12 Å)/AlOx and reported the observation of different excitation modes such as LF modes (<2 GHz), IF modes (2-8 GHz) and HF modes (12-18 GHz) in the perpendicular orientation, where the CoFeB thickness was 12 Å and the iDMI value was 1.2 mJ/m$^2$ ($D_s$ = 1.44 pJ/m)[18]. We estimated the iDMI strength of Re/Co (12 Å)/Pt equal to 2.47 mJ/m$^2$ ($D_s$ = 2.97 pJ/m). Here, we find a different behavior of resonant modes, where the LF mode is in the range of 2-9 GHz and the HF mode is in the range of $f$ = 10-25 GHz (Fig. 4(a)). From the above discussion on the Re/Co/Pt systems, it is clearly evident that with increasing the thickness, the strength of the iDMI decreases. This in turn decreases the start and end frequencies for the HF mode and LF modes in Re/Co/Pt system. Consequently, by adjusting the iDMI strength, it is feasible to tailor the frequency range for different resonant modes and the frequency range for the excitation modes depending on the different material parameters[17,18]. Further, our micromagnetic simulations of sample S4 reveal the existence of skyrmion core resonant modes in the frequency range of 0.5 to 8.5 GHz. In contrast, the Ta/CoFeB (12 Å)/AlOx exhibits the skyrmion core resonant modes in a higher frequency range of 12-15 GHz, called HF modes[18].

Both Ta/CoFeB (12 Å)/AlOx[18] and Re/Co (16 Å)/Pt systems exhibit IP easy axis magnetization, however they differ in terms of thickness, anisotropy and iDMI strength. That's why the skyrmion core resonant modes occur in different frequency ranges in these systems. Here, it is worth noting that skyrmion resonant mode are observed in Ir/Fe/Co/Pt at the frequency of 7-8 GHz, which occurs in LF mode region. This difference in observed skyrmion core resonant modes highlights how material composition and structural factors significantly impact the dynamic behavior of magnetic systems.

It is reported that by increasing the Co thickness in the multilayers, the stabilization of a skyrmion lattice is observed at larger field ranges[19] in the perpendicular configuration. As a consequence, the different resonant modes are easily distinguished (Figs. 4(g)-(i)). However, the skyrmions stabilizing in lower thickness Co (8 Å) exhibit a narrow field range for their formation, see Fig. 4(a). Therefore, only HF modes are observed in the higher frequency range of $f$ = 30-35 GHz as shown in the Fig.4(a) and other resonant modes are very weak and difficult to visualize.

Similarly, in IP configuration, the HF, IM, and LF modes due to labyrinth domains are clearly observed with a higher amplitude for 8 and 10 Å thick Co. Similar modes are observed in the simulated FMR spectra for the S2 sample as shown in Figs. 8(b, h, i & j). However, we assigned distinct names to three resonant modes (H0$_{DE}$, LF$_{DV}$ & HF$_{DV}$). We observed that, with increasing the thickness of Co to 12 or 16 Å, the LF and IF resonant modes vanish with creation of the ZF modes at zero magnetic field. We also found the similar type of modes called zero-field domain edge modes (H0$_{DE}$) in the simulated FMR spectra for sample S4 as shown in Figs. 8(d & o). The LF$_{DV}$ and HF$_{DV}$ modes are localized in the domain wall and volume of domains, respectively, due to presence of antiparallel upward and downward labyrinth chiral domains confirmed by the micromagnetic simulation (see Figs. 9(b & f)). As a result, it is possible to calculate the frequency band-gap between the LF$_{DV}$ and HF$_{DV}$ modes at zero field. The group velocity of the magnons can be estimated, as it is directly proportional to frequency band-gap at zero field. It is worth noting that a higher group velocity at zero field is advantageous for the magnon-based spintronics devices[25]. For IP configuration, one can describe the vanishing of the resonant modes due to the labyrinth domains by the magnetostatic interaction and iDMI strength for higher thicknesses of Co, i.e.20 and 24 Å. Here it is to be noted that the iDMI strength decreases with increasing the thickness. Also, the interlayer coupling can affect the resonant modes of parallel stripes. Furthermore, the resonant frequency modes due to labyrinth domains are vanished for 20 Å and 24 Å Co. Note that the effective damping (see Table 2) which depends on the Co thickness and spin pumping can also affect the resonant modes of the skyrmions and the labyrinth domains[17,18,19].

**Table 2. List of magnetic parameters from FMR measurement.** Magnetic parameters describing the [Re (10 Å)/Co(d$_{Co}$)/Pt(10 Å)]$_{20}$ multilayers extracted from FMR measurement.

| Samples | $\mu_0 M_{eff}$ (T) | $K_U$ (kJ/m$^3$) | $\frac{K_{4\perp}}{M_S}$ (mT) | $g$ | $\alpha^\perp$ | $\alpha^\parallel$ |
|---|---|---|---|---|---|---|
| S1 | -0.764 | 1724 | -12 | 2.08 |  | 0.03 |
| S2 | -0.395 | 1682 | -28 | 2.06 | 0.052 | 0.033 |
| S3 | 0.012 | 1295 | -16 | 2.15 | 0.025 | 0.037 |
| S4 | 0.168 | 1172 | -44 | 2.2 | 0.033 | 0.047 |
| S5 | 0.467 | 977 | -21 | 2.19 | 0.025 | 0.036 |
| S6 | 0.556 | 922 | -6 | 2.17 | 0.019 | 0.036 |

In this section, we analyze the damping from the FMR linewidth for both IP and OOP orientations (see sup. Fig. SM-2). We have observed that the damping decreases as the Co thickness increases, which is attributed to the higher spin pumping for lower Co thicknesses. For the IP orientation, the damping $\alpha^{\|}$ is higher than the damping $\alpha^{\perp}$ measured in OOP directions (except for sample S2), which can be explained in terms of two magnon scattering. We determined the anisotropy and effective magnetization by using the resonance equations for a tetragonally distorted cubic (001) system, i.e., distinguishing between IP and OOP anisotropy contributions. The effective magnetization ($\mu_0 M_{\text{eff}} = \mu_0 M_S - \frac{2K_{4\perp}}{M_S}$)[36,37] is negative for Co thicknesses of 8 Å and 10 Å. However, the $\mu_0 M_{\text{eff}}$ is positive for Co thicknesses of 16 Å, 20 Å, and 24 Å, which confirms the IP anisotropy. The 12 Å Co sample is at the spin reorientation transition, as FMR and SQUID show opposing results regarding the easy axis. However, the polar angular dependence measured by FMR reveals that the easy axis direction is tilted to 70° and not 90° (see supplementary Figs. SM-3(c & g)). Here, it should be noted that we have considered $M_{\text{eff}} < 0$ and $M_{\text{eff}} > 0$ for OOP and IP easy axis (for FMR measurements), respectively for our convenience.

**Summary**:

The static and dynamics properties of epitaxial Re/Co/Pt system were studied. During magnetization reversal, we observed the transition from labyrinth domains to skyrmions by field-dependent MFM. This transition is mainly influenced by the Co thickness and eventually the effective anisotropy. Specifically, with increasing Co thickness and decreasing effective anisotropy, we noticed a shift from isolated skyrmions to a skyrmion lattice in perpendicular configuration ($\theta_H$=90°). Furthermore, we investigate the stabilization of a skyrmion lattice at remanence by applying a static field under an inclined angle. Our findings demonstrate the tunability of the skyrmion density depending on both the effective anisotropy as well as the iDMI strength. By performing the angle-dependent measurements, we observed that the stabilization angle for a maximum skyrmion density shifts towards the lower angle ($\theta_H \approx 0°$). In dynamical characterization, we revealed the occurrence of different distinct resonant modes corresponding to the skyrmion and parallel stripe domain states from FMR measurements. The observed frequency range of the different resonant modes narrows as the iDMI strength decreases i.e., with increasing Co thickness. Additionally, we observed that the thicker Co layers shows more resonant modes in OOP configuration, whereas the thinner Co layers exhibit a higher number of resonant modes in IP configuration. Furthermore, from angle-dependent FMR measurements at $\theta_H$=78° and 30° for samples S2 and S4, respectively, we confirmed the frequencies range of resonant modes lying between those observed in IP and OOP configurations. To support the experimental results, we performed the micromagnetic simulation, and observed the different skyrmion-resonant modes such as $SK_C$, $SK_{GA}$, and bulk resonant modes such as $LF_B$ and $HF_B$ in OOP orientation. Additionally, we observed zero-field domain edge resonant mode $H0_{DE}$ and volume resonant modes such as $LF_{DV}$, and $HF_{DV}$ due to chiral stripes in IP orientation. The $SK_C$ and $SK_{GA}$ modes are prominent in Co (16 Å), while only $SK_{GA}$ is observed in Co (8 Å), highlighting their dependence on thickness of Co, anisotropy and iDMI strength. The formation of chiral stripes by applying IP field exhibits the different eigen modes and zero-field frequency band-gap, which could be useful for magnonic devices because of their higher group velocity. Finally, we determined the effective Gilbert damping for all samples showing lower in OOP configuration than in IP configuration. Moreover, we found that the damping value decreases as the Co thickness increases, which is attributed to a lower spin pumping. Our investigations on dynamics of epitaxial samples with significantly enhanced iDMI strength holds a great potential for future spintronics devices. The tunability of the resonant frequency through iDMI strength opens up new exciting possibilities to design and develop the GHz frequency-based electronics.

**Experimental and numerical methods:**

We fabricated the $Al_2O_3(0001)/Pt$ (30 Å)/[Re (10 Å)/Co ($d_{Co}$)/Pt (10 Å)]$_{20}$ multilayers by molecular beam epitaxy at a pressure of $3\times10^{-8}$ mbar. We prepared samples with Co thicknesses $d_{Co}$ of 8 Å, 10 Å, 12 Å, 16 Å, 20 Å, and 24 Å, respectively. The 30-Å-Pt buffer layer was deposited at 750 °C after annealing the $Al_2O_3(0001)$ substrate at 850 °C. The 20-Å-thick Pt capping layer prevents oxidation. All sample details are listed in table 3 (Materials and Method).

We determined the static magnetic parameters by superconducting quantum interference (SQUID) magnetometry. We performed the MFM imaging by taking the phase mode signal at 100 nm lift after the topography scan. We used low moment PPP-LM-MFMR tips from Nanosensors. For the angle-dependent MFM imaging, we first applied the tilted field to the sample by an external electromagnet and then performed the MFM imaging at remanence. For the field dependent MFM in perpendicular field, we used a permanent magnet. We set the field strength by the distance between sample and permanent magnet. The field of view of all MFM images is 5 μm × 5 μm.

Furthermore, we measured the frequency vs. field resonance dependence by vector network analyzer FMR in a frequency range of 0.1 to 35 GHz with sweeping field at fixed frequency using a coplanar waveguide (CPW) with an 80-μm-wide center conductor. We took the FMR spectra both in field-IP and OOP orientation.

**Table 3. List of the fabricated samples.**

| Sample stack | Sample name |
|---|---|
| $Al_2O_3(0001)/Pt$ (30 Å)/[Re (10 Å)/Co (8 Å)/Pt (10 Å)]$_{20}$/Pt (20 Å) | S1 |
| $Al_2O_3(0001)/Pt$ (30 Å)/[Re (10 Å)/Co (10 Å)/Pt (10 Å)]$_{20}$/Pt (20 Å) | S2 |
| $Al_2O_3(0001)/Pt$ (30 Å)/[Re (10 Å)/Co (12 Å)/Pt (10 Å)]$_{20}$/Pt (20 Å) | S3 |
| $Al_2O_3(0001)/Pt$ (30 Å)/[Re (10 Å)/Co (16 Å)/Pt (10 Å)]$_{20}$/Pt (20 Å) | S4 |
| $Al_2O_3(0001)/Pt$ (30 Å)/[Re (10 Å)/Co (20 Å)/Pt (10 Å)]$_{20}$/Pt (20 Å) | S5 |
| $Al_2O_3(0001)/Pt$ (30 Å)/[Re (10 Å)/Co (24 Å)/Pt (10 Å)]$_{20}$/Pt (20 Å) | S6 |

**Protocol for applying tilted field:**

First, the sample is saturated by applying a DC magnetic field with amplitude equal to the saturation field in negative direction. After reaching the saturation point, the field gradually is increased to reach the positive saturation point under a particular angle. Again, after reaching the positive saturation point at that particular angle, the field is gradually reduced back to zero. The MFM images are recorded finally at remanence. We repeat this procedure for every desired angle, while keeping the same negative saturation DC magnetic field so that the remanence state should be the same for every angle before applying the positive field. The amplitudes for positive saturation DC field are different for different angles, which have been extrapolated by fitting the straight line from perpendicular saturation field ($\theta_H = 0°$) to IP ($\theta_H = 90°$) saturation field measured by SQUID. The extrapolated values of positive saturation field $\mu_0 H_{+DC,Sat}$ are 1.1 T, 0.85 T, 0.55 T, 0.55 T, 0.65 T, and 0.75 T for samples S1, S2, S3, S4, S5, and S6, respectively.

**Micromagnetic simulations:**

For micromagnetic simulations, we use own our version of Mumax3 [32-33], called Amumax, which solves the Landau–Lifshitz–Gilbert equation:

$$\frac{dm}{dt} = \frac{\gamma\mu_0}{1+\alpha^2}(m \times H_{eff} + \alpha m \times (m \times H_{eff})),$$

where $m = M/M_S$ is the normalized magnetization, $H_{eff}$ is the effective magnetic field acting on the magnetization, $\gamma$ is the gyromagnetic ratio, $\mu_0$ is the vacuum permeability. The following components were considered in the effective magnetic field $H_{eff}$: demagnetizing field $H_d$, exchange field $H_{exch}$, uniaxial magnetic anisotropy field $H_{anis}$ and external magnetic field $H_{ext}$. Thermal effects were neglected. Thus, the effective field is expressed as:

$$H_{eff} = H_d + H_{exch} + H_{dmi} + H_{ext} + H_{anis} + h_{mf}.$$

where the last term, $h_{mf}$ is a microwave magnetic field used for SW excitation. The exchange and anisotropy fields are defined as

$$H_{exch} = \frac{2A_{ex}}{\mu_0 M_S}\Delta m,$$

$$H_{anis} = \frac{2K_{u,bulk}}{\mu_0 M_S} m_z \hat{z},$$

$$H_{dmi} = \frac{2D}{M_{sat}}\left(\frac{\partial m_z}{\partial x}, \frac{\partial m_z}{\partial y}, -\frac{\partial m_x}{\partial x} - \frac{\partial m_y}{\partial y}\right)$$

where $A_{ex}$ is the exchange constant, D is interfacial DMI constant.

To gain better insight into the static and dynamic properties of [Re(10Å)/Co($d_{Co}$)/Pt(10Å)]$_{20}$ multilayers, we modelled samples S2 and S4 with $d_{Co}$=10Å and 16Å. We simulated the complete 20-layer repetition of the Re/Co/Pt trilayer (with periodic boundary conditions in the film plane) to get the most accurate account of dipolar effects possible, since it is known that the skyrmion core deviates from the usual tubular structure to more complex three-dimensional configurations due to inhomogeneous dipolar fields along the multilayer thickness [30]. We used the following magnetic parameters: $M_s$ = 1.3374 MA/m and $K_u$ = 1.756 MJ/m³, $A_{ex}$=2 pJ/m, $iDMI$=0.003 J/m², and $M_s$ = 1.3374 MA/m and $K_u$ = 1.169 MJ/m³, $A_{ex}$=2.5 pJ/m, $iDMI$=0.00218 J/m² for samples S2 and S4, respectively. We used the gyromagnetic ratio as a fitting parameter to simulate the spin wave dynamics yielding $\gamma$=1.3095 Mrad/Tsreproducing closely the KM mode frequencies (see Figs. 8 (a-d)). For all calculations, we used a supercell size of 2 μm × 2 μm × 58 nm, which we discretized into 486 × 486 × 58 cells for sample S2 and 2 μm × 2 μm × 72 nm, which we discretized into 486 × 486 × 38 cells for S4. The simulated stack consists of 20 ferromagnetic layers, each 1.0 nm and 1.6 thick for S2 and S4 respectively, with 2.0 and 1.6 nm non-magnetic spacers between them. This was like the experimental setup, where the total thickness of non-magnetic Pt and Re layers separating adjacent ReCoPt layers is 2.0 nm. Periodic boundary conditions were applied in the x and y directions within the plane of the film. To simulate the pinning effects and nonuniformity observed in experimental samples, we employed Voronoi tessellation to create 20 nm grains, each with a unique value of perpendicular magnetic anisotropy. The anisotropy values within these grains were assigned based on a normal distribution cantered around the nominal value, with a standard deviation of 5%.

*Relaxation procedure:*
The relaxation simulations of the hybridized system were performed using a sequence of Amumax built-in relaxation functions in four steps: minimization (*minimize*), relaxation

(*relax*) with Gilbert damping α = 1.0 (for Co layers) and default energy threshold values, solving the full LLG equation (simulation run for 1 ns), and energy minimization again. This sequence allowed us to obtain a stable magnetization configuration.

In our simulations of the hysteresis loops–to avoid the overestimation of coercive field during the remagnetization of a uniformly saturated sample–we developed a "spin injection" method. This approach involves creating several dozens of 20 nm random grains (using Voronoi tessellation) with magnetization randomly oriented within the plane during the initial steps of decreasing the magnetic field in the hysteresis loop simulation. This process is repeated only until the grains are stabilized (i.e., they do not remagnetize back to the saturated state). In other words, for various initial field values, we artificially generate "nucleation sites" for remagnetization and attempt to relax and maintain them. This procedure is repeated until the remagnetized regions of the simulated structure are "locked in." We set a threshold for repeating the spin injection at 98% of total saturation. The application of this algorithm enabled us to relax partial remagnetization of the system, allowing for a more accurate reproduction of the hysteresis loop during remagnetization along the easy axis (OOP). To further assist nucleation, before last relaxation procedure, we also run the simulations at $Temp = 320K$ for 2 ns using a Gilbert damping parameter of α=0.1.

*Dynamic simulations:*

The SW spectral response in dependence on the external magnetic field directed along the *x,y*-axis were calculated for values from 2 T to 0 T in decrements of 100 mT. To speed up micromagnetic simulations, negative field values were copied from positive values, neglecting the effects of DMI interactions. To prevent numerical artifacts arising from a super-symmetry of the spins, we tip the external field by 0.0001 degrees from the *z*-axis and 0.0001 in *x*-axis. During the simulations, we first relax the magnetization in the system until we reach the ground state for specific field amplitude and angle. We then excite the SWs with a global microwave magnetic field along the perpendicular to the external magnetic field, uniform in space, with a *sinc* temporal profile, a cut-off frequency of 60 GHz, and a peak amplitude of $5 \times 10^{-3}$ T. The excitation field is applied for 1 ns, and we sample the magnetization dynamics at intervals of 2.66 ps over a period of 5 ns. To acquire the SW spectrum, we took the space- and time-resolved IP magnetization and applied a Hamming window along the time axis. We then computed the real discrete Fourier transform using the fast Fourier transform (FFT) algorithm along the time axis for each cuboid composing the system

To generate the mode visualizations, we individually processed each cuboid in the system by calculating the FFT of the x- and y-components of the magnetization over time. For a selected frequency, the magnitude of the complex number was mapped as color intensity. This process was repeated for each cuboid in the system to create the complete visualization.

**Acknowledgement**:

SKJ acknowledges to IEEE magnetic society student seed funding 2020-2021. This work was supported by the Foundation for Polish Science (FNP) under the European Regional Development Fund – Program [REINTEGRATION 2017 OPIE 14-20] and by the Polish National Science Centre projects: [2016/23/G/ST3/04196] and [2020/37/B/ST5/02299]. MZ acknowledges support from [2020/39/D/ST3/02378]. AW would like to acknowledge the M-ERA.NET 3 (2022/04/Y/ST5/00164).

**Author contributions**:

SKJ conceives and leads the project with the help of AW and EM. SKJ and AL have performed the SQUID and its analysis. SKJ and AP deposited the sample. SKJ performed all MFM images. SKJ, JL and KL have performed VNA-FMR and analysis of FMR results. PA and AP have performed the PMOKE for SQUID. MZ and MM performed the micromagnetic simulation and its analysis. SKJ wrote the manuscript. All authors contributed, revised the manuscript and discussed the results.

**Conflict of interest**

The authors declare that they have no conflict of interest.